# Nonvolatile reconfigurable polarization rotator at datacom wavelengths based on a $Sb_2Se_3$/Si waveguide

JORGE PARRA[1], MIROSLAVNA KOVYLINA[1], AMADEU GRIOL[1], AND PABLO SANCHIS[1,*]

[1]*Nanophotonics Technology Center, Universitat Politècnica de València, Camino de Vera s/n, 46022 Valencia, Spain*
**pabsanki@ntc.upv.es*

**Abstract:** Silicon photonics has become a key platform for photonic integrated circuits (PICs) due to its high refractive index and compatibility with complementary metal-oxide-semiconductor manufacturing. However, the inherent birefringence in silicon waveguides requires efficient polarization management. Here, we report a reconfigurable polarization rotator (PR) using a $Sb_2Se_3$/Si waveguide operating at datacom wavelengths (1310 nm), providing nonvolatile switching with zero static power consumption. The polarization conversion relies on the interference of hybrid electric-magnetic (EH) modes, which can be reconfigured by changing the $Sb_2Se_3$ state between amorphous and crystalline. Our experimental device exhibits a polarization conversion efficiency (PCE) and a polarization extinction ratio (PER) as high as -0.08 dB and 17.65 dB, respectively, in a compact footprint of just 21 µm length. Therefore, the proposed reconfigurable PR offers a compact and energy-efficient solution for polarization management in silicon photonics, with potential applications in data communication networks and emerging applications benefiting from polarization information encodings, such as optical neural networks and quantum computing.

## 1. Introduction

Silicon (Si) photonics has been established during the last decade as the mainstay platform for developing photonic integrated circuits (PICs) [1]. Silicon has unique advantages, including its high refractive index contrast, which allows for the miniaturization of optical components, and its compatibility with standard complementary metal-oxide-semiconductor (CMOS) manufacturing processes, enabling cost-effective mass production. As a result, silicon PICs have been successfully developed for high-speed interconnections in data centers and a variety of sensing applications [2–4]. The prospects of silicon PICs for advancements in computing and quantum applications have also gained significant interest in recent years [5,6]. However, one of the main advantages of silicon—the high refractive index—may also give rise to a strong birefringence. This birefringence generates a polarization-dependent behavior in photonic waveguides that may hinder the design and performance of silicon devices [7]. Therefore, the control of the polarization and the possibility of implementing polarization-diversity schemes are necessary for many applications [8], making polarization rotators (PRs) vital building blocks. PR devices enable the rotation between the two orthogonal polarizations, transverse-electric (TE) and transverse-magnetic (TM), typically supported by silicon waveguides. One of the more compact solutions is based on breaking the cross-section symmetry of the waveguide, which hybridized the propagation modes, thus allowing the optical power to be periodically transferred between the TE and TM polarizations [9–11]. However, such PR devices are usually passive, hampering the reconfigurability of the polarization state. Tunable PR devices have been reported, but they are based on the power-hungry silicon thermo-optic effect [12,13].

Chalcogenide-based phase-change materials (PCMs) are currently excellent candidates for enabling reconfigurability in silicon photonic devices with ultra-low power consumption due to their unique property to switch between nonvolatile states with a large change of the refractive index [14]. Furthermore, their integration into large-scale silicon photonics platforms has been recently demonstrated [15,16]. Among the available chalcogenides, antimony selenide ($Sb_2Se_3$) and antimony sulfide ($Sb_2S_3$) provide a change of the real part of the refractive index with negligible optical losses at telecom and datacom wavelengths [17]. Therefore, there is a strong interest in developing these materials to implement ultra-compact phase shifters with zero-static power consumption [18]. Nevertheless, such property could also be exploited to enable reconfigurability in other devices, such as mode converters [19] or microring resonators [20]. Recently, reconfigurable PRs with performance at 1550-nm optical wavelengths have been proposed through numerical simulation by adding $Sb_2Se_3$ in silicon photonic waveguides with breaking symmetry [21,22].

In this work, we experimentally demonstrate a reconfigurable PR in the silicon photonics platform using the ultralow-loss $Sb_2Se_3$ PCM in the O-band. By changing the state of the PCM, our PR can maintain the input polarization or switch to the orthogonal with a non-volatile response, i.e., zero holding power.

## 2. Working principle

The proposed PR consists of an asymmetric $Sb_2Se_3$/Si waveguide. The device concept and its working principle are illustrated in **Fig. 1(a)**. Asymmetry is achieved using a narrow silicon wire and depositing a thin $Sb_2Se_3$ layer on top and sideways. Polarization conversion is based on the interference of two hybrid electric-magnetic (EH) modes supported by the $Sb_2Se_3$/Si waveguide [9]. Reconfigurability is enabled by the refractive index change of the $Sb_2Se_3$ layer between the amorphous and crystalline states. The length of the PR is chosen to achieve an orthogonal polarization conversion between the amorphous and crystalline states. Hence, the input polarization is maintained at the device's output for the amorphous state, while polarization conversion to the orthogonal optical mode occurs at the output for the crystalline state.

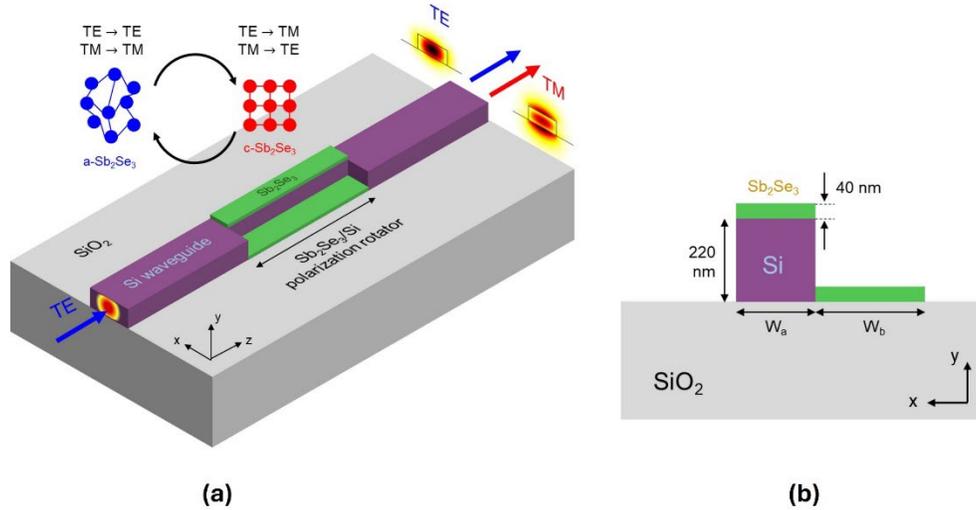

**Fig. 1.** Illustration of the proposed reconfigurable and nonvolatile polarization rotator based on an asymmetric $Sb_2Se_3$/Si waveguide. **(a)** 3D view of the device and illustration of the working operation when the fundamental TE mode is launched. In the amorphous state, the polarization is maintained at the output, whereas in the crystalline state, the polarization is rotated to TM. **(b)** Cross-section of the polarization rotator. The width of the device ($W_a + W_b$) is fixed at 500 nm. The upper cladding is air.

The cross-section of the considered hybrid waveguide is shown in **Fig. 1(b)**. The design of the reconfigurable PR starts from a standard 500 nm × 220 nm silicon-on-insulator (SOI) waveguide that is connected to a $Sb_2Se_3$/Si waveguide with a 40-nm-thick $Sb_2Se_3$ layer. The upper cladding is air. For the sake of simplicity in the device design, the width of the hybrid waveguide is set at 500 nm, and the device performance is optimized by varying the relation between the widths of the $Sb_2Se_3$/Si wire ($W_a$) and $Sb_2Se_3$ layer placed sideways ($W_b$).

## 3. Device design and optical performance

We obtained the EH modes supported by the hybrid waveguide and their effective refractive indices, $n_{eff}$, as a function of the value of $W_a$. The optical modes were calculated using an eigenmode solver based on finite element method (FEM) (see **Appendix A.1** for simulation details). Based on the values of $n_{eff}$, we calculated the length required to achieve a $\pi$ phase shift between the two EH modes, $L_\pi$, which gives rise to an orthogonal polarization conversion at the output. The value of $L_\pi$ is given by:

$$L_\pi = \frac{\lambda}{2|n_{\text{eff},1} - n_{\text{eff},2}|}, \tag{1}$$

where $\lambda$ is the working wavelength (1310 nm), and $n_{eff,1}$ and $n_{eff,2}$ are the effective refractive indices of the hybrid EH modes. The resulting values of $L_\pi$ for the amorphous and crystalline state as a function of $W_a$ and the $L_\pi$ ratio between both states are shown in **Fig. 2(a)**. The $Sb_2Se_3$/Si strips with a value of $W_a$ below 200 nm give both poor optical confinement and hybridization, thus yielding ineffective interference with high insertion loss. On the other hand, as the value of $W_a$ approximates 500 nm, the optical modes in the hybrid waveguide are less hybridized and become quasi-TE and -TM, turning off the polarization rotating performance. Consequently, based on these results, we choose $W_a$ = 250 nm as the optimal value due to the high confinement and mode hybridization for both amorphous and crystalline states [**Fig. 2(b)**]. Moreover, for $W_a$ = 250 nm, the ratio of $L_\pi$ between the amorphous and crystalline state is a rational number. This feature allows the selection of a device length, $L$, exhibiting opposite interference behavior between the amorphous and crystalline state, thus providing reconfigurable polarization conversion. The normalized polarization conversion, $PC$, can be calculated as:

$$PC = \sin^2\left(\frac{\pi}{2}\frac{L}{L_\pi}\right) \tag{2}$$

and it is shown for the amorphous and crystalline state in **Fig. 2(c)**. Based on these results, polarization reconfigurability with maximum polarization conversion difference between both states is achieved for L = 21 µm.

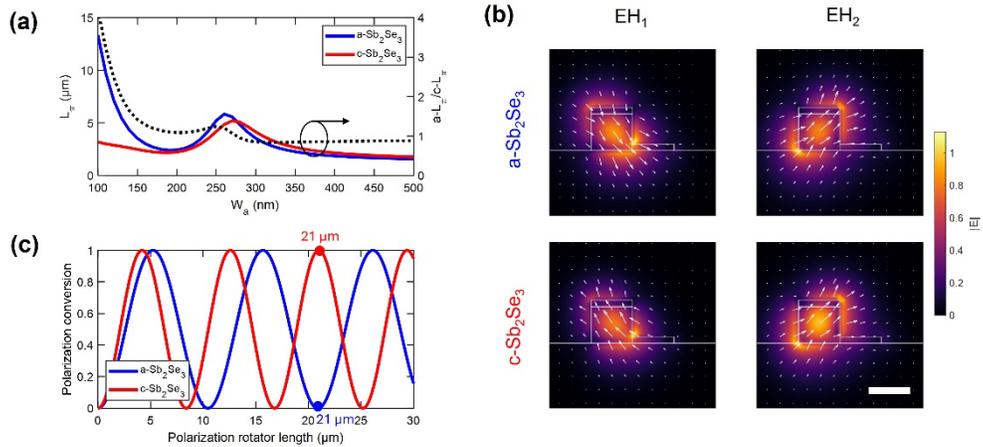

**Fig. 2. (a)** Length of the polarization rotator required to achieve a $\pi$ phase shift between the hybrid modes, $L_\pi$, (left y-axis) as a function of the value of the $W_a$ parameter for the amorphous

and crystalline states, and its ratio (right y-axis). **(b)** Electric field mode intensity profiles, |E|, and their field direction for $W_a$ = 250 nm. The scale bar is 250 nm. **(c)** Normalized polarization conversion as a function of the device length for $W_a$ = 250 nm and the amorphous and crystalline states. All results are given at $\lambda$ = 1310 nm.

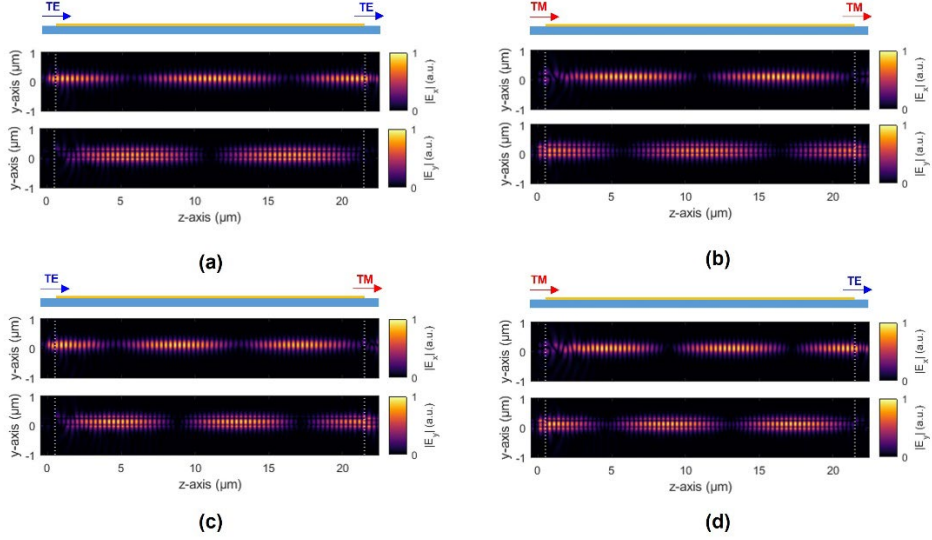

**Fig. 3.** $|E_x|$ and $|E_y|$ field distribution along the longitudinal cross-section of the optimal PR device for different polarization inputs and $Sb_2Se_3$ states. **(a)** TE launch and amorphous. **(b)** TM launch and amorphous. **(c)** TE launch and crystalline. **(d)** TM launch and crystalline. All results are given at $\lambda$ = 1310 nm.

Then, we obtained the performance of the optimal PR using 3D finite-difference time-domain (3D-FDTD) simulations (Methods). **Figure 3** shows the simulated electric field distribution of the optimal 21-µm-long PR device when launching the TE and TM optical modes and for the amorphous and crystalline states. In the amorphous state [**Figs. 3(a)** and **3(b)**], we verify that the length of the PR is two times the beat length ($2L_\pi$) of the hybrid modes, thereby maintaining the polarization state at the output. On the other hand, for the crystalline state [**Fig. 3(c)** and **3(d)**], the effective optical path traveled by the hybrid EH modes is increased by $L_\pi$, thus producing a 90º rotation of the polarization at the output of the PR device.

Based on the optimal design, we analyzed the impact on the optical performance of the PR caused by variations in the thickness and refractive index of the $Sb_2Se_3$ layer. We carried out 3D-FDTD simulations at 1310 nm to obtain the transmission of the PR in the amorphous and crystalline states, and for the TE and TM polarization when the TE mode was launched (**Figure 4**). By considering only variations on the $Sb_2Se_3$ thickness, the PR keeps its reconfigurable response with a tolerance of ±6 nm in both states although at expense of a lower polarization extinction ratio [**Figs. 4(a) and 4(b)**]. On the other hand, by fixing the $Sb_2Se_3$ thickness and varying its refractive index, the PR would exhibit a more robust performance against variations in the $Sb_2Se_3$ refractive index [**Figs. 4(c) and 4(d)**] compared to deviations in the thickness.

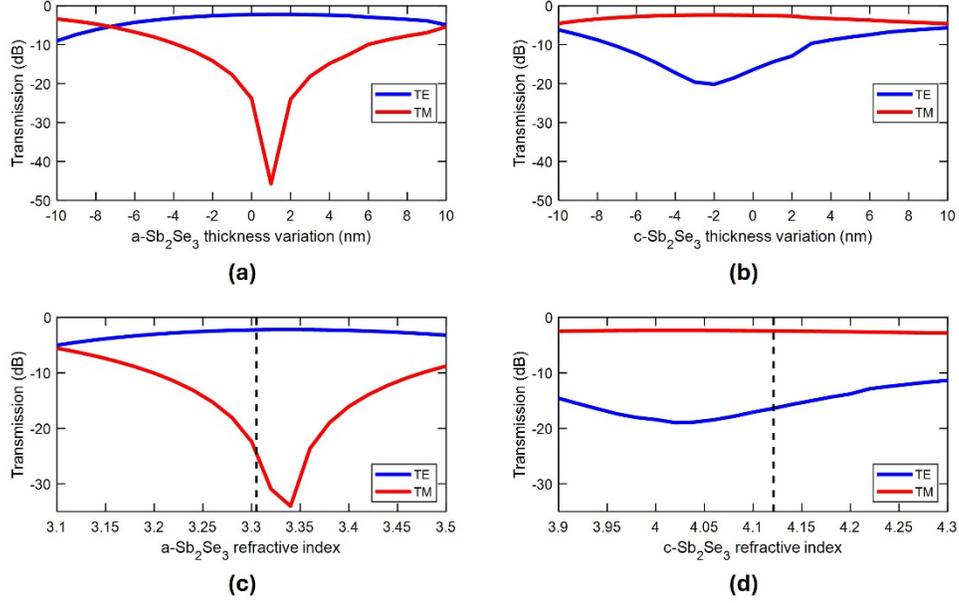

**Fig. 4.** Tolerance analysis of the optimal PR design against variations in the thickness and refractive index of the $Sb_2Se_3$ layer. Results show the transmission of the PR in the amorphous and crystalline states for the TE and TM polarization when the TE mode is launched at 1310 nm. (a,b) $Sb_2Se_3$ thickness variation with respect to 40 nm in the (a) amorphous and (b) crystalline states. (c,d) $Sb_2Se_3$ refractive index change in the (c) amorphous and (d) crystalline states. The dashed lines stand for the considered refractive index in design.

The simulated spectral performance of the optical device is shown in **Fig. 5** for the TE mode as input. A very similar performance was obtained by launching the TM mode. We calculated the polarization conversion efficiency (PCE) as:

$$PCE = \frac{P_{out}^{TE/TM}}{P_{out}^{TE} + P_{out}^{TM}} \qquad (3)$$

where $P_{out}^{TE}$ and $P_{out}^{TM}$ is the output power contained in the TE and TM polarization, respectively, and the value of the numerator is chosen for the desired output polarization. On the other hand, the polarization extinction ratio (PER) was calculated as the ratio between the desired and undesired polarization, i.e.:

$$PER = \frac{P_{out}^{TM/TE}}{P_{out}^{TE/TM}}. \qquad (4)$$

Based on our simulations, near-perfect PCE [**Fig. 5(a)**], together with PER values above 20 dB [**Fig. 5(b)**], are achieved at 1310 nm. The device's insertion loss (IL) at 1310 nm is below 3 dB for both states [**Fig. 5(c)**], and it is caused by the optical mismatch between the mode profiles of the fundamental mode of the Si waveguide and the hybrid modes excited in the $Sb_2Se_3$/Si waveguide. On the other hand, as the PR working principle is based on an interferometric effect, the bandwidth of the PR is affected by the dispersion in the effective refractive index of the hybrid modes as well as their field profile distribution, i.e., the amount of E-field in the x and y components, which also suffer wavelength dependence.

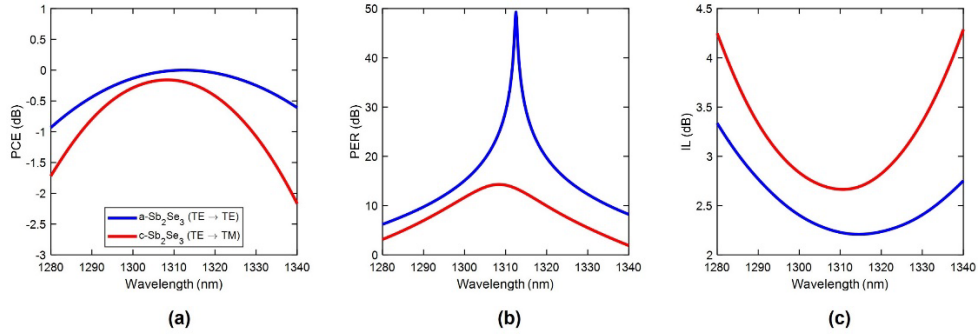

**Fig. 5.** Simulated spectral performance of the optimal PR device when the TE mode is launched for the amorphous (a-$Sb_2Se_3$) and crystalline (c-$Sb_2Se_3$) states. **(a)** Polarization conversion efficiency (PCE). **(b)** Polarization extinction ratio (PER). **(c)** Insertion loss (IL). Results are shown for the optimal PR and using 3D-FDTD simulations.

## 4. Fabrication and experimental results

We fabricated and characterized the proposed PR device on a standard 220-nm-thick SOI photonic chip (see **Appendix A.2** for fabrication details). To determine the experimental performance of the fabricated PR, we used a PIC comprised of the PR device, a polarization beam splitter (PBS), and TE- and TM-optimized grating couplers to collect separately the power contained in each polarization [**Fig. 6(a)**]. The values of $W_a$ and $W_b$ in the fabricated device were 280 nm and 235 nm [**Fig. 6(b)**]. Before injecting the optical power into the PR device, we polarized the light to TE using an external polarization controller and a TE reference waveguide (see **Appendix A.3** for characterization setup details). On the other hand, the PR device was first measured in the as-deposited amorphous state and then in the crystalline state by heating the chip at 250 ºC for 10 min. in an Ar atmosphere.

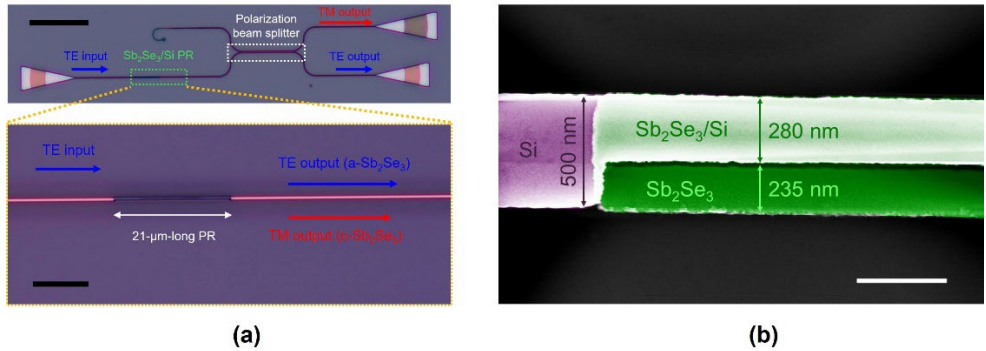

**Fig. 6. (a)** Optical micrograph of the PIC used to characterize the optical performance of the proposed PR device. The top and bottom scale bars are 50 μm and 10 μm, respectively. **(b)** False-color scanning electron microscope (SEM) image of the fabricated PR device. The scale bar is 400 nm.

The measured transmission of the PR under TE polarization at the input for the amorphous and crystalline states is shown in **Fig. 7**. We obtained the transmission of the PR by subtracting the spectral response of the grating couplers and PBS that were characterized from reference photonic structures as described in **Appendix B**. The experimental performance of the PR in the amorphous state shows a good agreement with the expected simulated response [**Fig. 7(a)**]. The TM output increased its optical power when the PR was switched to the crystalline state. In contrast, the TE output suffered a notable decrease caused by the polarization rotation of the optical mode at the output of the PR device [**Fig. 7(b)**]. Discrepancies between simulations and experimental measurements could be attributed to the interplay of several factors such as variations in the actual refractive index and thickness of the $Sb_2Se_3$ as discussed in section 3 (see **Fig. 4**), as well as the slight deviation in the $W_a$ and $W_b$ values of the fabricated PR compared to the target values [see **Fig. 6(b)**].

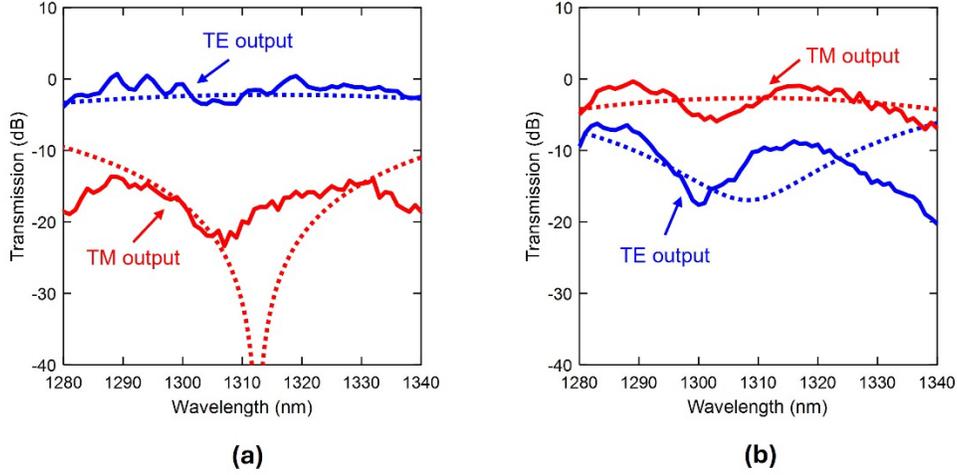

**Fig. 7.** Experimental (solid lines) and simulated (dotted lines) spectral response of the PR device under TE polarization at the input in the **(a)** amorphous and **(b)** crystalline states.

Based on the measured spectral response, we calculated the PCE and PER of the fabricated PR shown in **Figs. 8(a)** and **8(b)**, respectively. At 1310 nm, our PR exhibits a PCE value higher than -1 dB in both states and a PER of almost 20 dB for the amorphous state. A moderate PER of ~5 dB is found in the crystalline state.

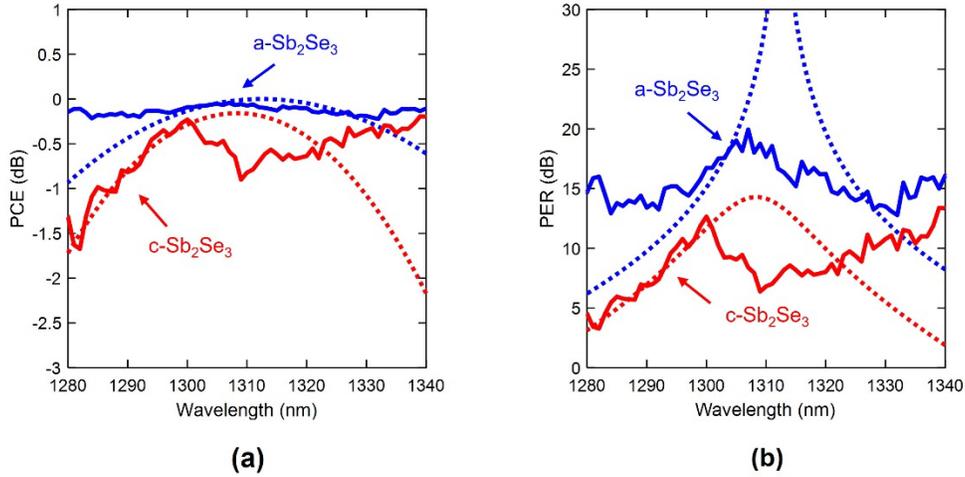

**Fig. 8.** Experimental (solid lines) and simulated (dotted lines) performance of the PR device in the amorphous (a-$Sb_2Se_3$) and crystalline (c-$Sb_2Se_3$) states. **(a)** Polarization conversion efficiency (PCE). **(b)** Polarization extinction ratio (PER).

## 5. Discussion and conclusions

In this work, we have reported a reconfigurable PR device in the SOI platform working at 1310 nm with a nonvolatile response, i.e., with zero static power consumption to hold the reconfigurable states. Our device is based on a hybrid $Sb_2Se_3$/Si waveguide. It can maintain the input polarization of the fundamental optical mode or switch to the orthogonal polarization by changing the $Sb_2Se_3$ material state between amorphous and crystalline, respectively. On the other hand, our 21-µm-long fabricated device features a PER as high as 10 dB when the polarization is rotated ($Sb_2Se_3$ in the crystalline state) with a low IL of 2.2 dB at 1310 nm. The IL of the device could be reduced in future work by optimizing the interface between the Si and $Sb_2Se_3$/Si waveguide by offsetting the $Sb_2Se_3$/Si in the x-axis to provide better overlapping and using a taper structure. According to our simulations, introducing a -100 nm displacement in the x-axis [Fig. 1(a)] would yield a reduction of the IL in both states, achieving values as low as -0.84 dB and -1 dB for the amorphous and crystalline state, respectively, at 1310 nm. On the other hand, on-chip thermo-electrical switching of our device could be implemented using transparent heaters based on graphene [23,24] or transparent conductive oxides [25–27], as well as a doped silicon-based heaters [16–18,20]. In the latter, it should be noticed that the design of the PR would rely on an asymmetric doped-Si slab waveguide to enable Joule heating. In addition, the relatively long length of the $Sb_2Se_3$ section may complicate the heater design due to the challenge of cooling quickly enough for an effective amorphization. In this case, a segmented approach could be considered for a more efficient dissipation of the heat [16,18].

Compared to previous silicon-based polarization management devices working at 1310 nm (**Table 1**), our PR device exhibits good optical performance compared to passive devices with a drastic improvement in terms of footprint and power consumption compared to large and power-hungry PR devices based on the silicon thermo-optic effect. The optical bandwidth could be increased by means of an adiabatic mode evolution-based rotation approach, usually achieved with asymmetric directional couplers [28], but at the expense of a larger footprint.

Table 1. Comparison of experimental silicon-based polarization control devices working in the O-band.

| Ref. | Type | IL (dB) | PER (dB) | Footprint | Bandwidth | Reconfigurable | Static power consumption |
|---|---|---|---|---|---|---|---|
| [28] | PSR | <1 | >15 | 1.6 µm × 85 µm | 80 nm | No | - |
| [29] | PSR | <1 | >20 | 3 µm ×17 µm | 50 nm | No | - |
| [30] | PBS | <2.5 | >10 | 28 µm × 20 µm | 80 nm | No | - |
| [31] | PBS | <1 | >10 | 15 µm × 30 µm | >100 nm | No | - |
| [32] | PR | <1.5 | >10 | 300 µm × 500 µm | 60 nm | Yes (Heaters) | 18 mW |
| This work | PR | <2.5 | >5 | 0.5 µm × 21 µm | 40 nm | Yes ($Sb_2Se_3$) | Zero |

PSR = Polarization splitter rotator; PBS = Polarization beam splitter; PR = Polarization rotator.

Finally, such a reconfigurable, compact, and ultralow-energy consumption PR device could serve as a key building block for photonic applications benefiting from harnessing the polarization state of light and leveraging the scalability of PICs, such as datacom networks [3], optical neural networks [33,34], or quantum computing [35,36].

## APPENDIX A: METHODS

### 1. Optical simulations

We considered the following materials' refractive index values at 1310 nm for performing optical simulations: $n_{air}$ = 1, $n_{SiO2}$ = 1.444, $n_{Si}$ = 3.5, $n_{a\text{-}Sb2Se3}$ = 3.305 (amorphous) [15], $n_{c\text{-}Sb2Se3}$ = 4.121 + j0.001 (crystalline) [15]. We employed a FEM eigenmode solver [FemSIM tool from RSoft] to calculate the optical mode profiles and their effective refractive indices. We used a nonuniform mesh consisting of a 30 nm x 30 nm bulk grid with minimum divisions of 10 points in both the x and y axes. The simulated domain was 2 µm × 2 µm. The optical performance of the polarization rotator was simulated by means of 3D-FDTD using the FullWAVE tool from RSoft. In the XY plane, we used the same simulation settings as for FEM simulations. In the z-axis (propagation direction), we used a 30 nm bulk grid. A perfectly matched layer (PML) consisting of 10 PML cells was employed as the boundary condition. An overlap monitor with the TE and TM fundamental mode of the Si waveguide was placed at the device's output to determine the optical power contained in each polarization.

### 2. Fabrication

The integrated photonic devices were fabricated starting from a diced SOI wafer of 220 nm thick Si layer on a 3 µm thick $SiO_2$ buried oxide layer (SEHE). The photonic devices were patterned using an electron-beam (e-beam) lithography tool (JBX-8100FS) using an accelerating voltage of 100 kV and 500 µC/cm² dose. A 300 nm thick negative tone resist (ma-N 2405) was exposed for creating the silicon waveguides, whereas 270 nm and 400 nm thick positive tone resists of polymethylmethacrylate (PMMA) were utilized for defining the trenches of the grating couplers and $Sb_2Se_3$ windows, respectively. The silicon layer was dry-etched based on inductively coupled plasma–reactive ion etching (ICP-RIE) (Corial 210L) using a mixture of $SF_6/C_4F_8$ gases. The remaining resist on the photonic devices was removed using oxygen plasma for 10 min (Tepla PV200). The $Sb_2Se_3$ layer was fabricated by e-beam evaporation (Pfeiffer Classic 500) at a base vacuum pressure of $2.4\times10^{-7}$ mbar and working pressure of $8\times10^{-7}$ mbar using 4 kV High Voltage and 1.5 mA current. The rate of the deposited layer was controlled by a water-cooled quartz crystal monitor to be 0.5 Å/s. A mechanical profilometer (KLA) was used to assess the final thickness of 40±4 nm.

### 3. Optical characterization setup

The integrated photonic structures were characterized by employing an optical fiber-based setup. A continuous-wave optical signal was generated by using a stepped external-cavity tunable laser (OSICS T100). The laser output power was set to 0 dBm, and the wavelength step employed was 1 nm. The light was injected/extracted to/from the PIC grating couplers using single-mode optical fibers titled 10º out of the plane of the chip. 3-axis motorized stages were employed for precise fiber-chip alignment. The light was polarized before being injected into the chip using a manual 3-paddle polarization controller (Thorlabs FPC032). Finally, the output light was collected using a high-sensitivity photodiode (Thorlabs S154C) connected to an optical power meter (Thorlabs PM320). All measurements were carried out at room temperature.

## APPENDIX B: REFERENCE PHOTONIC STRUCTURES

We fabricated reference photonic structures comprising TE and TM grating couplers connected by 250-µm-long Si waveguides. The spectral response of these reference photonic structures is shown in **Fig. 9**. The grating period was 500 nm and 780 nm for the TE and TM grating couplers, respectively. For both designs, the fill factor was 50 %, and the etching of the trenches was 70 nm.

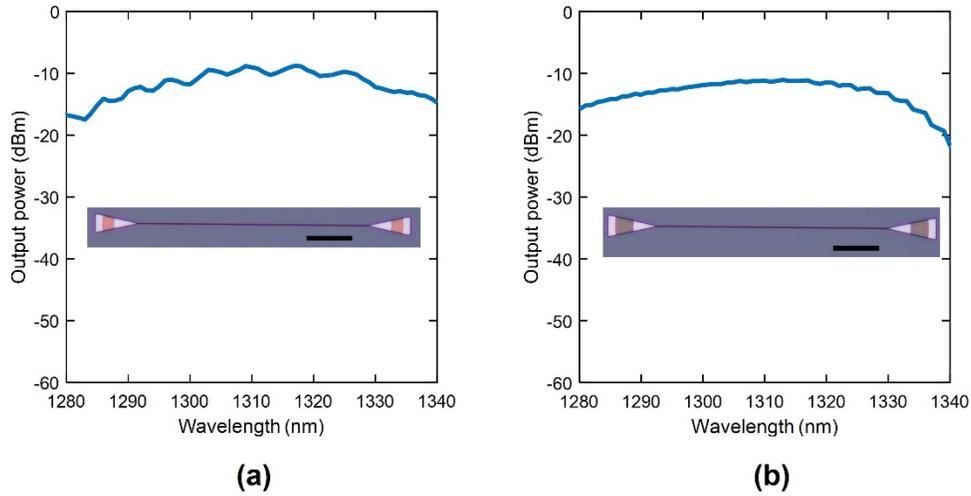

**Fig. 9.** Measured spectral response of reference waveguides using **(a)** TE- and **(b)** TM-designed grating couplers. The inset shows an optical micrograph of the tested reference photonic structures. The scale bar is 50 µm.

The employed PBS was based on a 41.6-µm-long symmetric directional coupler with a 300 nm gap between waveguides. The bending radius was 10 µm. The measured spectral response is shown in **Fig. 10** under a TE and TM polarized light. We measured the PBS using TE [**Fig. 10(a)**] and TM [**Fig. 10(b)**] grating couplers at both input and output.

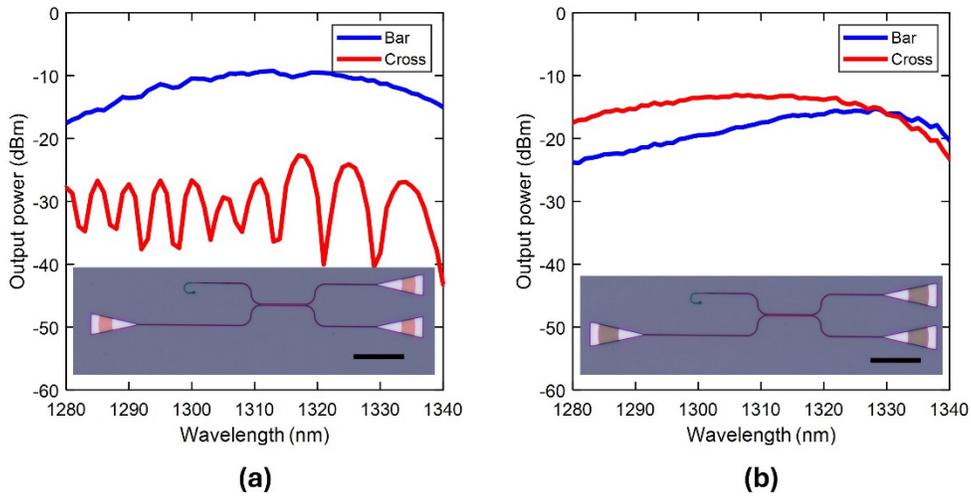

**Fig. 10.** Measured spectral response of the PBS structure for the **(a)** TE and **(b)** TM polarization. The top images show an optical micrograph of the measured structures. The scale bar is 50 µm.

**Funding.** Agencia Estatal de Investigación (TED2021-132211B-I00, PID2022-137787OB-I00); Generalitat Valenciana (PROMETEO Program (CIPROM/2022/14)); Universitat Politècnica de València (Grants PAID-06-23, PAID-10-23).

**Acknowledgments.** The authors would like to thank the clean room staff of the Nanophotonics Technology Center for the fabrication of the device and T. Angelova for her help with the characterization of the $Sb_2Se_3$.

**Disclosures.** The authors declare no conflict of interest.

**Data availability.** Data underlying the results presented in this paper are not publicly available at this time but may be obtained from the authors upon reasonable request.